\documentclass[12pt,showpacs]{revtex4}
\usepackage[english]{babel}
\usepackage{graphicx}
\usepackage{epsfig}

\newcommand{\bea}{\begin{eqnarray}}
\newcommand{\eea}{\end{eqnarray}}
\newcommand{\beq}{\begin{equation}}
\newcommand{\eeq}{\end{equation}}

\newcommand{\nn}{\nonumber}
\def\dis{\displaystyle}

\begin{document}

\def\dsdt{$\frac{d\sigma}{dt}$}


\def\See{\vec e^{\prime *}\cdot \vec e\,}
\def\Sss{\vec s^{\prime *}\cdot \vec s\,}
\def\Ses{\vec e^{\prime *}\cdot \vec s\,}
\def\Sse{\vec s^{\prime *}\cdot \vec e\,}

\def\Sek{\vec e^{\prime *}\cdot \hat{\vec k}\,}
\def\Ske{\vec e \cdot \hat{\vec k}'\,}

\def\Ssk{\vec s^{\prime *}\cdot \hat{\vec k}\,}
\def\Sks{\vec s \cdot \hat{\vec k}'\,}

\def\Skk{\hat{\vec k}'\cdot \hat{\vec k}\,}


\def\Sk{\vec\sigma \cdot \hat{\vec k} \,}
\def\kS{\vec\sigma \cdot \hat{\vec k}'\,}

\def\Se{\vec\sigma \cdot \vec e \,}
\def\eS{\vec\sigma \cdot \vec e'\,}
\def\Ss{\vec\sigma \cdot \vec s \,}
\def\sS{\vec\sigma \cdot \vec s'\,}


\def\Vee{\vec\sigma \cdot \vec e^{\prime *} \times \vec e\,}
\def\Vss{\vec\sigma \cdot \vec s^{\prime *} \times \vec s\,}

\def\Vkk{\vec\sigma \cdot \hat{\vec k}' \times \hat{\vec k}\,}

\def\Vek{\vec\sigma \cdot \vec e^{\prime *} \times \hat{\vec k}\,}
\def\Vke{\vec\sigma \cdot \vec e \times \hat{\vec k}'\,}

\def\dis{\displaystyle}

\title{Proton spin polarizabilities from polarized Compton scattering}
\normalsize
\author{B. Pasquini}
\affiliation{
Dipartimento di Fisica Nucleare e Teorica, Universit\`{a} di Pavia, and\\
Istituto Nazionale di Fisica Nucleare, Sezione di Pavia, I-27100 Pavia, Italy}
\author{D. Drechsel}
\affiliation{ Institut f\"ur Kernphysik, Johannes Gutenberg Universit\"at,
D-55099 Mainz, Germany}
\author{M. Vanderhaeghen}
\affiliation{
Physics Department, College of William and Mary, Williamsburg, VA 23187, USA, and\\
Theory Center, Thomas Jefferson National Accelerator Facility,
Newport News, VA 23606, USA}

\begin{abstract}
Polarized Compton scattering off the proton is studied within the
framework of subtracted dispersion relations for photon energies up
to 300 MeV. As a guideline for forthcoming experiments, we focus the
attention on the role of the proton's spin polarizabilities and
investigate the most favorable conditions to extract them 
with a minimum of model dependence. We
conclude that a complete separation of the four spin polarizabilities
is possible, at photon energies between threshold 
and the $\Delta(1232)$ region, provided one can achieve polarization 
measurements with an accuracy of a few percent.
\end{abstract}


\date{\today}
\maketitle
\section{Introduction}

The polarizabilities of a composite system such as the nucleon are
elementary structure constants, just as its size and shape. They can
be studied by applying electromagnetic fields to the system. The
physical content of the nucleon polarizabilities can be visualized
best by effective multipole interactions for the coupling of the
electric $(\vec{E})$ and magnetic $(\vec{H})$ fields of a photon
with the internal structure of the nucleon. This structure can be
accessed experimentally by the Compton scattering process $\gamma +
N \to \gamma + N$ on the nucleon, see
Refs.~\cite{DPV_report,Schumacher} for recent reviews. When
expanding the Compton scattering amplitude in the energy of the
photon, the zeroth and first order terms follow from a low energy
theorem and can be expressed solely in terms of the charge, mass,
and anomalous magnetic moment of the nucleon. The second order terms
in the photon energy describe the response of the nucleon's internal
structure to an electric or magnetic dipole field, they are given by
the following effective interaction~:
\begin{eqnarray}
\label{h2order}
H_{\rm{eff}}^{(2)} = -4\pi \left[
{\textstyle\frac{1}{2}}\,\alpha_{E1}\,\vec{E}\,^2 +
{\textstyle\frac{1}{2}}\,\beta_{M1}\,\vec{H}\,^2\ \right],
\end{eqnarray}
where the proportionality coefficients are the electric
($\alpha_{E1})$ and magnetic ($\beta_{M1}$) dipole (scalar) polarizabilities,
respectively. These global structure coefficients are proportional
to the electric and magnetic dipole moments of the nucleon which are
induced when placing the nucleon in static electric and magnetic
fields. They have been measured extensively using unpolarized
Compton scattering. A fit to all modern low-energy Compton
scattering data yields the following results for the
proton~\cite{Olmos01}:
\begin{eqnarray}
\alpha_{E1}^p & = & \left[ 12.1  \pm
0.3(\rm{stat}) \mp 0.4(\rm{syst}) \pm 0.3(\rm{mod}) \right]
\times 10^{-4} \, \mbox{fm}^3 \ ,
\nonumber \\
\beta_{M1}^p & = & \left[ 1.6 \pm 0.4(\rm{stat}) \pm 0.4(\rm{syst}) \pm
0.4(\rm{mod}) \right] \times 10^{-4} \, \mbox{fm}^3 \ ,
\label{eq:exp}
\end{eqnarray}
with the statistical, systematical and model-dependent errors, respectively. 
These values confirm, beyond any doubt, the dominance of the
electric polarizability $\alpha_{E1}$. The tiny value of the
magnetic polarizability, $\beta_{M1}$, comes about due to a
cancellation of the large paramagnetic contribution of the $N \to
\Delta$ spin-flip transition with a nearly equally large 
diamagnetic contribution, partly due to pion loop effects.
\newline
\indent The internal spin structure of the nucleon appears at third
order in an expansion of the Compton scattering amplitude. It is
described by the effective interaction
\begin{eqnarray}
\label{h3order}
{H}_{\rm{eff}}^{(3)} &=&  - 4 \pi \left[
{\textstyle\frac{1}{2}}\gamma_{E1E1}
\,\vec{\sigma}\cdot(\vec{E}\times\dot{\vec{E}})
+ {\textstyle\frac{1}{2}}\gamma_{M1M1}\,
\vec{\sigma}\cdot(\vec{H}\times\dot{\vec{H}}) \right. \nonumber \\
&&\left. \hspace{.75cm}
- \gamma_{M1E2}\, E_{ij}\,\sigma_iH_j
+ \gamma_{E1M2}\, H_{ij}\,\sigma_iE_j\  \right],
\end{eqnarray}
which involves one derivative of the fields with regard to either
time or space~: $\dot{\vec{E}} = \partial_t\vec{E}$ and
$E_{ij}=\frac{1}{2}(\nabla_iE_j+\nabla_jE_i)$, respectively. The
four spin (or vector) polarizabilities $\gamma_{E1E1}$, $\gamma_{M1M1}$,
$\gamma_{M1E2}$, and $\gamma_{E1M2}$ describing the nucleon spin response 
at third order, can be related to a multipole
expansion~\cite{BGM}, as is reflected in the subscript notation. For 
example, $\gamma_{M1E2}$ corresponds to the excitation of the
nucleon by an electric quadrupole ($E2$) field and its de-excitation by a
magnetic dipole ($M1$) field. Expanding the Compton scattering amplitude to
higher orders in the energy, one obtains higher order
polarizabilities to the respective order, e.g., the quadrupole
polarizabilities to fourth order~\cite{BGM,HDPV}.
\newline
\indent The effective Hamiltonians of Eqs.~(\ref{h2order}) and
(\ref{h3order}) describe a shift in the nucleon energies at second
order in the electromagnetic fields. This has been implemented in
recent years as a tool to calculate nucleon polarizabilities in
lattice QCD. By calculating the mass shifts in a constant background
field, and isolating the quadratic response, it has thus been
possible to compute the electric polarizability of the neutron and
other neutral octet and decuplet baryons~\cite{Christensen:2004ca},
and the magnetic polarizability of the proton,  neutron, and all
other particles in the lowest baryon octet and decuplet
states~\cite{Lee:2005dq}. In Ref.~\cite{Lee:2005dq}, 
$\beta_{M1}^p$ has been calculated by use of the 
Wilson action in the pion mass range
$0.5 \leq m_\pi \leq 1$~GeV and neglecting disconnected loop diagrams. 
For the smallest calculated pion mass
of $m_\pi \simeq 500$~MeV, a value of $\beta_{M1}^p = ( 2.36 \pm
1.20 ) \times 10^{-4} \, \mbox{fm}^3$ was obtained. While it is
encouraging to see that the lattice result is in the right ballpark
when compared to the experimental value of Eq.~(\ref{eq:exp}), a
more precise comparison clearly requires a dynamical fermion
calculation, including disconnected loop diagrams and much smaller
pion masses. Such small pion masses would then allow one to
extrapolate safely to the physical pion mass within the framework of
chiral perturbation theory.
\newline
\indent To determine the spin polarizabilities, Eq.~(\ref{h3order})
can likewise be used to calculate energy shifts of a polarized
nucleon in an external field. As an example, consider a nucleon
polarized along the $z$-axis and apply a magnetic field rotating
with angular frequency $\omega$ in the $xy$ plane,
\begin{eqnarray}
\vec H = B_0 \left[ \cos (\omega t) \; \hat e_x + \sin (\omega t) \; \hat e_y
\right],
\end{eqnarray}
where $\hat e_i$ stands for the unit vector in the direction $i = x,
y$ and $B_0$ is the magnitude of the field. Such a field leads to
an energy shift $\Delta E = \mp 2 \pi \gamma_{M1 M1} \omega B_0^2$
if the nucleon spin is oriented along the positive ($-$) or negative
($+$) $z$-axis. The split between the two levels is then directly
proportional to the magnetic dipole spin polarizability
$\gamma_{M1M1}$.
\newline
\indent It has been shown in
Ref.~\cite{Detmold:2006vu} that allowing for
background fields with suitable variations in space and time,
lattice QCD should be able to calculate all six dipole
polarizabilities. In particular, calculations are in progress to
determine the electric polarizability of a charged particle such as
the proton as well as the four proton spin polarizabilities of
Eq.~(\ref{h3order})~\cite{Detmold:2006xh}.
\newline
\indent 
A microscopic understanding of the nucleon's
polarizabilities requires to quantify the interplay between
resonance contributions, e.g., the $N \to \Delta$ transition, and
long range pion cloud effects. Such systematic studies of the pion
cloud effects became possible with the development of chiral
perturbation theory (ChPT), by systematically expanding the
lagrangian in the external momenta and the pion mass
(``$p$-expansion''). The first such calculation to the one-loop order,
at $\mathcal{O}(p^3)$, yielded the following leading terms for the
proton scalar polarizabilities~\cite{Bernard}:
\begin{eqnarray}
\alpha_{E1}^p = 10\beta_{M1}^p = \frac
{5\alpha_{em}g_A^2}{96\pi f_{\pi}^2m_{\pi}} = 12.2\ ,
\label{eq:lo}
\end{eqnarray}
where $\alpha_{em} = 1/137$, $g_A \simeq 1.27$, and $f_\pi =
92.4$~MeV. This result is in remarkable agreement with the
experimental result of Eq.~(\ref{eq:exp}). It also illustrates that
these quantities diverge in the chiral limit, which is a challenge
for the lattice QCD calculations. Conversely, if one is in the
``small $m_\pi$'' regime where the chiral expansion converges well,
ChPT can complement the lattice calculations by extrapolating to the
physical pion mass. The ChPT work was extended to $\mathcal{O}(p^4)$
within the heavy-baryon expansion~\cite{BKSM}. However, the
agreement with the experimental values of the scalar
polarizabilities became much more challenging when including the
$\Delta$ degree of freedom, which yields a large
contribution to $\beta_{M1}^p$~\cite{Hemmert,HHKK,Pascalutsa:2004wm}.  
\newline
\indent The spin polarizabilities have been calculated within the
framework of ChPT at $\mathcal{O}(p^3)$~\cite{BKKM} and
$\mathcal{O}(p^4)$~\cite{Ji:1999sv,VijayaKumar:2000pv,Gellas:2000mx}.
However, much less is known about these observables on the
experimental side, except for the forward ($\gamma_0$) and backward
($\gamma_{\pi}$) spin polarizabilities of the proton, given
by the following linear combinations of the polarizabilities of
Eq.~(\ref{h3order}):
\begin{eqnarray}
\gamma_{0}&=&-\gamma_{E1E1}-\gamma_{M1M1}-\gamma_{E1M2}-\gamma_{M1E2}
\,,\label{eq:g0_mult}\\
\gamma_{\pi}&=&-\gamma_{E1E1}+\gamma_{M1M1}-\gamma_{E1M2}+\gamma_{M1E2}
\,.\label{eq:gpi_mult}
\end{eqnarray}
The forward spin polarizability has been determined from the 
Gerasimov-Drell-Hearn sum rule experiments at MAMI and
ELSA~\cite{GDHmami,GDHelsa},
\begin{eqnarray}
\label{eq:go} \gamma_0=( -1.00\pm 0.08\pm 0.10)\times 10^{-4} \:
\mbox{fm}^4\,,
\end{eqnarray}
and the recent experimental value for the backward spin
polarizability has been obtained by a dispersive analysis of
backward angle Compton scattering~\cite{Schumacher},
\begin{eqnarray}
\gamma_\pi=(-38.7 \pm
1.8)\times 10^{-4} \: \mbox{fm}^4\;.\label{eq:gpi}
\end{eqnarray}
No data exist for the other two independent proton spin
polarizabilities. It is the aim of the present work to
show that polarized Compton scattering, both near pion threshold and
in the $\Delta(1232)$ resonance region, can be used to
determine the remaining two spin polarizabilities.
\newline
\indent The extraction of polarizabilities from Compton scattering
data has been performed by three techniques. The first one is a low
energy expansion of the Compton cross sections. Unfortunately this
procedure is only applicable at photon energies well below 100~MeV,
which makes a precise extraction a rather challenging task. The
sensitivity to the polarizabilities is increased by measuring
Compton scattering observables around pion threshold and into the
$\Delta(1232)$ resonance region. 
A second formalism which has been successfully applied to Compton data 
in this energy region makes use of dispersion relations. 
This formalism has been worked out for both unsubtracted~\cite{lvov97} and
subtracted~\cite{DGPV} dispersion relations, and yield the values  
of the scalar polarizabilities given in Eq.~(\ref{eq:exp}).  
Recently, a third approach has been
developed within the framework of a chiral effective field
theory~\cite{McGovern,Pascalutsa,Hildebrandt_unp,Hildebrandt_pol}. 
For energies below and around pion threshold
the full Compton scattering process has been calculated to fourth
order in the small momenta, allowing for an independent extraction
of the polarizabilities from Compton scattering data. The thus
obtained values for $\alpha_{E1}^p$ and $\beta_{M1}^p$ 
in the work of \cite{McGovern,Hildebrandt_unp} are nicely
compatible with the results given by Eq.~(\ref{eq:exp}).
\newline
\indent In this work we study the polarized proton Compton
scattering observables both below pion threshold and in the
$\Delta(1232)$ resonance region within the subtracted dispersion
relation formalism of Ref.~\cite{DGPV}. In particular we investigate
the sensitivity of different beam and beam-target polarization
observables to the extraction of the proton spin polarizabilities.
\newline
\indent
After a brief review of fixed-$t$ dispersion relations for the Compton 
scattering process in Section 2, we discuss different single 
and double polarization observables in Section 3. We present our results 
for these different polarization observables in Section 4 and study the 
sensitivity to the extraction of the proton spin polarizabilities. 
Finally, we give our conclusions in Section 5.


\section{Fixed-t subtracted dispersion relations}
\label{disp}

In this section we review the essentials of the dispersion relation formalism
for real Compton scattering (RCS), a more detailed presentation can be found in
Refs.~\cite{DPV_report,DGPV}. Let us first define the kinematics of RCS on the
proton, the reaction \beq \gamma(q) + p(p) \rightarrow \gamma(q') + p(p'), \eeq
where  the variables in brackets denote the four-momenta of the participating
particles. The familiar Mandelstam variables are
\begin{equation}
s=(q+p)^2\ ,\ \ t=(q-q')^2\ ,\ \ u=(q-p')^2\ ,
\label{eq2.1}
\end{equation}
which are constrained by $s+t+u=2M^2$, where $M$ is the nucleon mass. The
crossing-symmetric variable $\nu$ is defined by
\begin{equation}
\nu=\frac{s-u}{4M}\, .
\label{eq2.3}
\end{equation}
The two Lorentz invariant variables $\nu$ and $t$ are related to the initial
($E_\gamma$) and final ($E'_\gamma$) photon lab energies and to the lab
scattering angle $\theta_{{\rm lab}}$  by
\begin{eqnarray}
\nu&=&E_\gamma+\frac{t}{4M}=\frac{1}{2}(E_\gamma+E'_\gamma),\nn\\
t&=&-4E_\gamma \, E'_\gamma \, \sin^2 (\theta_{\rm lab}/2)=
-2M(E_\gamma-E'_\gamma).\nn
\end{eqnarray}
The $T$ matrix of real Compton scattering can be expressed by 6 independent
structure functions $A_i(\nu, t)$, $i=1,...,6$, which were first introduced in
Ref.~\cite{lvov97}. These structure functions depend on $\nu$ and $t$, they are
free of kinematic singularities and constraints, and because of the crossing
symmetry they satisfy the relation $A_i(\nu, t)=A_i(-\nu, t)$. Assuming further
analyticity and an appropriate high-energy behavior, the amplitudes $A_i$
fulfill unsubtracted dispersion relations (DRs) at fixed $t$,
\begin{equation}
{\rm Re} A_i(\nu, t) \;=\; A_i^B(\nu, t) \;+\;
{2 \over \pi} \; {\mathcal P} \int_{\nu_{thr}}^{+ \infty} d\nu' \;
{{\nu' \; \mathrm{Im}_s A_i(\nu',t)} \over {\nu'^2 - \nu^2}}\;,
\label{eq:unsub}
\end{equation}
where $A_i^B$ are the nucleon pole contributions of the Born terms
describing the photon scattering off a point-like nucleon with
anomalous magnetic moment, as explicitly given in App.~A of
Ref.~\cite{lvov97}. Furthermore, $\mathrm{Im}_s A_i$ are the
discontinuities across the $s$-channel cut of the Compton process,
starting at pion production threshold, i.e.,  $\nu_{thr} = m_\pi +
(m_\pi^2 + t/2 )/(2 M)$, with $m_\pi$ the pion mass. However, as can
be deduced from the asymptotic behavior of the functions
$A_i(\nu,t)$  for $\nu \rightarrow \infty$ and fixed
$t$~\cite{DGPV}, such unsubtracted DRs do not converge for the
amplitudes $A_1$ and $A_2$. We therefore subtract the fixed-$t$ DRs of
Eq.~(\ref{eq:unsub}) at $\nu=0$, with the result
\begin{equation}
\mathrm{Re} A_i(\nu, t) \;=\; A_i^B(\nu, t) \;+\;
\left[ A_i(0, t) - A_i^B(0, t) \right]
\;+\;{2 \over \pi} \;\nu^2\; {\mathcal P} \int_{\nu_{thr}}^{+ \infty} d\nu' \;
{{\; \mathrm{Im}_s A_i(\nu',t)} \over {\nu' \; (\nu'^2 - \nu^2)}}\;.
\label{eq:sub}
\end{equation}
Because of the two additional powers of $\nu'$ in the denominator, these
subtracted DRs should now converge for all of the invariant amplitudes.
\newline
\indent The 6 subtraction functions $A_i(\nu=0, t)$ appearing in
Eq.~(\ref{eq:sub}) can be determined by once-subtracted DRs in the
variable $t$:
\begin{eqnarray}
A_i(0, t) \;-\; A_i^B(0, t) &=&
\left[ A_i(0, 0) \;-\; A_i^B(0, 0) \right]
\;+\;
\left[ A_i^{t-pole}(0, t) \;-\; A_i^{t-pole}(0, 0) \right] \nonumber\\
&+&\;{t \over \pi} \; \int_{4 \, m_\pi^2}^{+ \infty} dt' \; {{\mathrm{Im}_t
A_i(0,t')} \over {t' \; (t' - t)}} \;+\;{t \over \pi} \; \int_{-
\infty}^{-2m^2_\pi-4Mm_\pi} dt' \; {{\mathrm{Im}_t A_i(0,t')} \over {t' \; (t'
- t)}} \;, \label{eq:subt}
\end{eqnarray}
where $A_i^{t-pole}(0, t)$ represents the contribution of the poles in the $t$
channel, in particular of the $\pi^0$ pole in the case of $A_2$ as explicitly
evaluated in Ref.~\cite{DGPV}, and the subtraction constants $a_i = A_i(0, 0)
\;-\; A_i^B(0, 0)$ are related to the polarizabilities as explained below.
\newline
In order to evaluate the dispersion integrals in Eq.~(\ref{eq:sub}), the
imaginary parts in the $s-$channel are calculated from the unitarity relation,
taking into account the $\pi N$ intermediate states and the resonant
contributions of inelastic channels involving more pions in the intermediate
states. In particular, for the $\gamma N\rightarrow \pi N \rightarrow \gamma N$
contribution we use the multipole amplitudes from the analysis of Hanstein {\it
et al.}~\cite{HDT} for energies $\nu\leq 500 $ MeV, and at the higher energies
up to $\nu=1.5$ GeV we  take the solutions of the SAID analysis~\cite{said}.
The multi-pion intermediate states are approximated by the inelastic decay
channels of the $\pi N$ resonances as detailed in Ref.~\cite{DGPV}. This simple
approximation of the higher inelastic channels is quite sufficient, because
these channels are largely suppressed by the energy denominator
$\nu'(\nu'^2-\nu^2)$ in the subtracted DRs of Eq.~(\ref{eq:sub}).
\newline
The imaginary part in the $t$-channel integral from $4 m_\pi^2
\rightarrow + \infty$ in Eq.~(\ref{eq:subt}) is saturated by the
possible intermediate states for the $t$-channel process, which lead
to cuts along the positive $t$-axis. For values of $t$ below the $K
\bar K$ threshold, the $t$-channel discontinuity is dominated by the
two-pion intermediate states, $\gamma\gamma\rightarrow
\pi\pi\rightarrow N\bar N$. We calculate this contribution by
evaluating a unitarized amplitude for the $\gamma\gamma\rightarrow
\pi\pi$ subprocess, and then combine it with the $\pi \pi\rightarrow
N\bar N$ amplitudes as determined from dispersion theory by
analytical continuation of the $\pi N$ scattering
amplitudes~\cite{Hoehler83}. The second integral in
Eq.~(\ref{eq:subt}) extends from $-\infty$ to $ -2m^2_\pi
-4Mm_\pi\approx - 0.56$ GeV$^2.$ As long as we stay at small
(negative) values of $t$, this integral is strongly suppressed by
the denominator $t'(t'-t)$ in Eq.~(\ref{eq:subt}), and therefore it
can be approximated by the contributions of $\Delta$-resonance and
non-resonant $\pi N$ intermediate states. The latter contributions
are evaluated by first evaluating the imaginary parts of the Compton
amplitude in the physical $s$-channel region by unitarity, and then
extrapolating these results into the unphysical region at $\nu=0$
and negative $t$ by means of analytical continuation.
\newline
\indent The 6 subtraction constants $a_i$ in Eq.~(\ref{eq:subt}) are
related to the electric ($\alpha_{E1}$) and magnetic ($\beta_{M1}$)
scalar polarizabilities in the spin-independent sector,
\begin{eqnarray}
\label{eq:ai-spinind}
\alpha_{E1}=-\frac{1}{4\pi}(a_1+a_3+a_6)\,,\quad\quad\beta_{M1}=\frac{1}{4\pi}(a_1-a_3-a_6)\,,
\end{eqnarray}
and to the 4 spin-dependent or vector polarizabilities,
\begin{eqnarray}\label{eq:ai-spindep}
\gamma_{E1E1}&=&\frac{1}{8\pi M}(a_2-a_4+2 a_5+a_6), \qquad\qquad
\gamma_{M1M1}=-\frac{1}{8\pi M}(a_2+a_4+2a_5-a_6),
\nonumber\\
\gamma_{E1M2}&=&\frac{1}{8\pi M}(a_2-a_4-a_6),\qquad\qquad \qquad
\gamma_{M1E2}=-\frac{1}{8\pi M}(a_2+a_4+a_6)\,.
\end{eqnarray}

\indent Although in principle all 6 subtraction constants $a_1$ to $
a_6$ could be used as fit parameters, we restrict the fit to the
parameters $a_5$ and $a_6$, or equivalently to the two spin
polarizabilities $\gamma_{E1E1}$ and $\gamma_{M1M1}$. For fixed
values of the fit parameters $a_5$ and $a_6$, the subtraction
constants $a_1$ and $a_3$ are then determined by a recent global fit
of the scalar polarizabilities to the low-energy data~\cite{Olmos01},
\begin{eqnarray}\label{alpha-beta-exp}
\alpha_{E1}+\beta_{M1}&=&(13.8\pm0.4) \times 10^{-4} \, \mbox{fm}^3 \, ,
\nonumber\\
\alpha_{E1}-\beta_{M1}&=&(10.5\pm0.9\, (\mathrm{stat.+syst.})\pm0.7\,
(\mathrm{mod.})) \times 10^{-4} \, \mbox{fm}^3\,.
\end{eqnarray}
In the spin-dependent sector, we use the experimental values of the
forward ($\gamma_0$) and backward ($\gamma_\pi$) spin
polarizabilities  describing the Compton spin-flip amplitude at
$\theta=0^{\circ}$ and $180^{\circ}$, respectively. These
observables have been expressed in Eqs.~(\ref{eq:g0_mult}) and
(\ref{eq:gpi_mult}) by linear combinations of the polarizabilities
defined in Eq.~(\ref{eq:ai-spindep}), they are related to the
subtraction constants as follows:
\begin{eqnarray}
\gamma_{\pi}=-\frac{1}{2\pi M}(a_2+a_5) \,,\quad \quad \gamma_{0}=
\frac{1}{2\pi M}a_4. \label{eq:gammapi0}
\end{eqnarray}
In particular, the subtraction constant $a_2$ is fixed by the value
of $\gamma_\pi$ given by Eq.~(\ref{eq:gpi}). This definition of the
backward spin polarizability includes both the dispersive and the
large $\pi^0$-pole contribution. In the analysis of
Ref.~\cite{Schumacher}, the latter takes the value
$\gamma_\pi^{\pi^0-{\rm pole}}= -46.7 \times 10^{-4}$ fm$^4$, which
leads to $\gamma_\pi^{{\rm disp}}= (8.0\pm 1.8 )\times 10^{-4}$
fm$^4$. In fact, the pion pole contribution is not known to that
accuracy, and therefore the error in Eq.~(\ref{eq:gpi}) stems from
both the dispersive and the pole contributions to the backward spin
polarizability. The value of the forward spin polarizability, or
equivalently of $a_4$, is fixed by Eq.~(\ref{eq:go}). We conclude
that the choice of the subtraction constants $a_5$ and $a_6$ as fit
parameters is equivalent to varying the polarizabilities
$\gamma_{E1E1}$ and $\gamma_{M1M1}$. As can be seen from
Eqs.~(\ref{eq:ai-spindep}) and (\ref{eq:go}), another possibility
would be to fit $a_2$ and $a_6$ or $\gamma_{E1M2}$ and
$\gamma_{M1E2}$.

\section{Cross sections and asymmetries - formalism}
\label{formalism }

The general formalism for RCS with one or two polarized particles
has been originally derived in Ref.~\cite{BGM}. In the following we
only review some pertinent formulas necessary to define the
observables for polarized incident photons scattered by polarized
target nucleons.
\newline
\indent
We work in a reference frame with the $z$ axis along $\hat q$ (the direction of
the incoming photon), the $x$ axis in the scattering plane and in the
half-plane of the outgoing photon, and the $y$ axis perpendicular to the
scattering plane along the direction $(\hat q\times \hat q')$. The
photon-polarization  density matrix is defined by the Stokes parameters $\xi_i$
($i=1,2,3$) as follows~\cite{landau}:
\begin{eqnarray}\label{Stocks-def}
\langle \, \varepsilon_\alpha \varepsilon_\beta^* \, \rangle =
\frac12(1+\vec\sigma \cdot \vec\xi)_{\alpha\beta} =  \frac12 \pmatrix{ 1+\xi_3
& \xi_1 -i\xi_2 \cr \xi_1 +i\xi_2 & 1-\xi_3 }_{\alpha\beta} ~ ,
\end{eqnarray}
where $\varepsilon^\mu$ is the photon polarization vector chosen in
the radiation gauge, $ \vec \varepsilon\cdot\vec q=0$, and
$\alpha,\beta =1,2$ denote either of the two orthogonal directions
$x$ and $y.$ The total degree of photon polarization is given by
$\xi=\sqrt{\xi_1^2+\xi^2_2+\xi^2_3}$, and
$\xi_{\ell}=\sqrt{\xi^2_1+\xi_3^2}$ and $\xi_2$ describe the degrees of
linear and circular polarization, respectively. Furthermore,
$\xi_2=+1$ and $\xi_2=-1$ correspond to right- and left-handed
states, respectively, and in the case of linear polarization the
azimuthal angle $\phi$ between the electric field and the scattering
plane is defined by $\cos 2\phi=\xi_3/\xi_{\ell}$ and $\sin
2\phi=\xi_1/\xi_{\ell}$. With these definitions, the Stokes parameters
take the same value in the c.m. and lab frames.
\newline
\indent The nucleon polarization density matrix is described by a
polarization four-vector $S^\mu$ that is orthogonal to the nucleon
four-momentum \cite{landau},
\begin{eqnarray}
\langle \, u(p)\bar u(p) \, \rangle=\frac12\Big( \gamma \cdot p + M\Big)
\Big(1+\gamma_5\gamma \cdot S \Big),
\end{eqnarray}
where $u(p)$ is a nucleon Dirac spinor normalized as $\bar u(p)\,u(p)=2M$.
\newline
\indent 
The differential cross section is related to the $T$-matrix by
\begin{eqnarray}\label{cs-generic}
\frac{d\sigma}{d\Omega}= \Phi^2\left|T\right|^2, \qquad \mbox{with} \qquad \Phi
= \cases{  \dis \frac {1}{8 \pi M}\frac{E'_\gamma}{E_\gamma}  & (lab frame) \cr
\dis \frac {1}{8 \pi \sqrt{s}}  & (c.m. frame). \rule{0ex}{4ex}}
\end{eqnarray}
The $T$-matrix for polarized photons and polarized targets can be decomposed in
8 independent functions $W_{ij}$,
\begin{eqnarray}\label{TWgN}
\left|T(\roarrow\gamma \roarrow N \to \gamma  N)\right|^2 &=& W_{00} +
W_{03}\xi_3  + N \cdot S\,(W_{30} + W_{33}^+\xi_3) \nn\\&+& K\cdot S\,(
W_{11}^+\xi_1 + W_{12}^+\xi_2) +  Q \cdot S\,( W_{21}^+\xi_1 + W_{22}^+\xi_2),
\end{eqnarray}
with the orthogonal four-vectors $K$, $N$, and $Q$ defined as
\begin{eqnarray} K_{\mu}=\frac{1}{2}(q'+q)_{\mu}, \qquad
N_{\mu}=\epsilon_{\mu\alpha\beta\gamma}P'^\alpha Q^\beta K^\gamma, \qquad
Q_{\mu}=\frac12(p-p')_{\mu}=\frac12(q'-q)_{\mu},
\end{eqnarray}
where $\epsilon_{0123}=1$. The photon asymmetry $\Sigma$ follows if the nucleon
polarization vector is set to zero, and the unpolarized case $\left|T\right|^2 = W_{00}$ is
recovered for vanishing photon and nucleon polarization vectors.
\newline \indent
In terms of the invariant amplitudes $A_i$, the functions $W_{ij}$
read~\cite{frolov60, BGM}:
\begin{eqnarray}\label{Wij}
   W_{00}
   &=&  \frac14 (4 M^2 -t) \Big( t^2 |A_1|^2 +
   \eta^2 |A_3|^2\Big) - \frac14\Big( t^3 |A_2|^2 -
   \eta^3 |A_4|^2\Big)  \nn\\ && {}
   -\nu^2 t\,(t + 8 \nu^2)|A_5|^2
   +\frac12 \eta\, (t^2 + 2 M^2 \eta)|A_6|^2 \nn\\ && {}
   +{\rm Re}\Big\{2 \nu^2 t^2(A_1 + A_2)A_5^* +
   \frac12\eta^2 (4 M^2 A_3 + t A_4)A_6^*\Big\}, \\[2ex]
   W_{03}
   &=&  \frac{\eta t}{2} {\rm Re}\Big\{ \Big( (4 M^2 - t) A_1
    + 4 \nu^2 A_5 \Big) A_3^* \,+\, 4 M^2 A_1 A_6^* \Big\}, \\[2ex]
   W_{30}
   &=& - 8 \nu{\rm Im}(t A_1 A_5^* + \eta A_3 A_6^*), \\[2ex]
   W_{33}^{\,\pm}
   &=& {\rm Im} \Big\{ {-} 8\nu\, \Big[ \Big(
     t A_1 - (t+4\nu^2) A_5 \Big) A_6^*  + \eta A_3 A_5^* \Big]
  \nn\\ && {} \qquad
     \pm \frac2m (t A_2 - 4\nu^2 A_5)(\eta A_4^* + t A_6^*)\Big\}, \\[2ex]
   W_{11}^{\,\pm}
   &=& {\rm Im}\Big\{ \frac{t}{2M}
   \Big( (4M^2-t)A_1 + 4\nu^2 A_5\Big) (\eta A_4^* + t A_6^*)
  \nn\\ && \qquad {}
   \pm 2\nu t \, (t A_2 - 4\nu^2 A_5) A_6^* \Big\}, \\[2ex]
   W_{12}^{\,\pm}
         &=&  {\rm Re} \Big\{ {-} \frac{\eta}{2M}
     \Big( (4M^2-t) A_3 + 4M^2 A_6 \Big) (\eta A_4^* + t A_6^*)
  \nn\\ && \qquad {}
   \pm 2\nu t\, (t A_2 - 4\nu^2 A_5) A_5^*  \Big\}, \\[2ex]
   W_{21}^{\,\pm}
         &=& 2 {\rm Im}\Big\{   {-}M\,(t A_2 - 4\nu^2 A_5)
     \Big( \eta A_3^* + (t+4\nu^2) A_6^* \Big)  \nn\\ && \hspace{8em} {}
      \pm \nu \Big( t A_1 - (t+4\nu^2) A_5 \Big)
     (\eta A_4^* + t A_6^*) \Big\}, \\[2ex]
   W_{22}^{\,\pm}
           &=&  2 {\rm Re}\Big\{ {-}M t\,(t A_2 - 4 \nu^2 A_5) A_1^*
   \mp \nu \eta A_3\,(\eta A_4^* + t A_6^*) \Big\},
\end{eqnarray}
where we have introduced the invariant variable $\eta= 4\nu^2+t-t^2/(4M^2).$
Below the pion photoproduction threshold the functions $A_i$ are real, and
therefore only the 6 structures $W_{00}, W_{03}, W_{12}^\pm$, and $W_{22}^\pm$
contribute below threshold.
\newline \indent
In the following, we focus on the asymmetries that can be obtained by varying
the photon and target polarizations in Eq.~(\ref{TWgN}):
\begin{itemize}
\item{}
circular photon polarization ($\xi_2=\pm 1$) and
target spin aligned in the $\pm z$ direction,
\begin{eqnarray}
\Sigma_{2z}
= \frac{\sigma^R_{+z}-\sigma^L_{+z}}{\sigma^R_{+z}+\sigma^L_{+z}}
= \frac{\sigma^R_{+z}-\sigma^R_{-z}}{\sigma^R_{+z}+\sigma^R_{-z}}
=\frac{C^{K}_{z}W_{12}^+ + C^{Q}_{z}W_{22}^+}{W_{00}} \, ,
\end{eqnarray}
where the coefficients $C^{K,Q}_{z}$ can be written in terms of lab or
invariant variables as
\begin{eqnarray}\label{eq:cqkz}
C^{K}_z&=&-\frac12\,(E_\gamma + E'_\gamma\cos\theta_{{\rm lab}})
= -\frac{s-M^2}{2M} - \frac{t\,(s+M^2)}{4M\,(s-M^2)}, \nn \\
C^{Q}_z&=& \frac12\,(E_\gamma - E'_\gamma\cos\theta_{{\rm lab}})
= - \frac{t\,(s+M^2)}{4M\,(s-M^2)}\, ,
\end{eqnarray}
\item{}
circular photon polarization ($\xi_2=\pm 1$) and
target spin aligned in the $\pm x$ directions,
\begin{eqnarray}\label{Sigxz}
\Sigma_{2x}
= \frac{\sigma^R_{+x}-\sigma^L_{+x}}{\sigma^R_{+x}+\sigma^L_{+x}}
= \frac{\sigma^R_{+x}-\sigma^R_{-x}}{\sigma^R_{+x}+\sigma^R_{-x}}
= \frac{C^{K}_{x}W_{12}^+ + C^{Q}_{x}W_{22}^+}{W_{00}}   \, ,
\end{eqnarray}
with \begin{eqnarray}\label{eq:cqkx} C^{K}_x=
C^{Q}_x=-\frac{1}{2}\,E'_\gamma\sin\theta_{{\rm lab}}
= - \frac{M\sqrt{-\eta t}} {2(s-M^2)} \,,
\end{eqnarray}
\item{}
linearly polarized photons, either parallel or
perpendicular to the scattering plane $(\xi_3 = \pm 1)$, and target nucleon
polarized perpendicularly to the scattering plane,
\begin{eqnarray}\label{eq:cny}
\Sigma_{3y}
&=& \frac{(\sigma^\parallel - \sigma^\perp)_{y}
-(\sigma^\parallel - \sigma^\perp)_{-y}}
{(\sigma^\parallel + \sigma^\perp)_{y}
+(\sigma^\parallel + \sigma^\perp)_{-y}}
=C^{N}_{y}\frac{W_{33}^+}{W_{00}}\,,
\end{eqnarray}
with
\begin{eqnarray}
C^N_y=\frac{M}{2}E_\gamma E'_\gamma\sin\theta_{{\rm
lab}}=\frac{M}{4}\sqrt{-\eta t}\, ,
\end{eqnarray}
\item{}
linearly polarized photons, either parallel or
perpendicular to the scattering plane $(\xi_3 = \pm 1)$, and
unpolarized target nucleons,
\begin{eqnarray}\label{eq:Sigma}
\Sigma_3
&=& \frac{\sigma^\parallel - \sigma^\perp}
{\sigma^\parallel + \sigma^\perp}
=\frac{W_{03}}{W_{00}}\,,
\end{eqnarray}
\item{}
photons linearly polarized at $\varphi = \pm\pi/4$ with respect to the
scattering plane ($\xi_1 = \pm 1)$ and the
nucleon target polarized in the scattering plane in the $\pm z$ direction,
\begin{eqnarray}\label{Sig1xz}
\Sigma_{1z}
= \frac{\sigma^{\pi/4}_{+z}-\sigma^{-\pi/4}_{+z}}
{\sigma^{\pi/4}_z+\sigma^{-\pi/4}_z}
= \frac{\sigma^{\pi/4}_{+z}-\sigma^{\pi/4}_{-z}}
{\sigma^{\pi/4}_{+z}+\sigma^{\pi/4}_{-z}}
= \frac{C^{K}_{z}W_{11}^+ + C^{Q}_{z}W_{21}^+}{W_{00}}\, ,
\end{eqnarray}
with coefficients $C^{K,Q}_z$ as defined in Eqs.~(\ref{eq:cqkz}),
\item{}
photons linearly polarized at $\varphi =
\pm\pi/4$ with respect to the scattering plane ($\xi_1 = \pm 1)$ and nucleon
targets polarized in the $\pm x$ direction,
\begin{eqnarray}
\Sigma_{1x}
= \frac{\sigma^{\pi/4}_{+x}-\sigma^{-\pi/4}_{+x}}
{\sigma^{\pi/4}_{+x}+\sigma^{-\pi/4}_{+x}}
= \frac{\sigma^{\pi/4}_{+x}-\sigma^{\pi/4}_{-x}}
{\sigma^{\pi/4}_{+x}+\sigma^{\pi/4}_{-x}}
= \frac{C^{K}_{x}W_{11}^+ + C^{Q}_{x}W_{21}^+}{W_{00}}\; ,
\end{eqnarray}
with $C^{K,Q}_x$ given in Eq.~(\ref{eq:cqkx}).
\end{itemize}

\section{Results and discussion}
\label{results}

\begin{figure}[ht]
\begin{center}
\epsfig{file=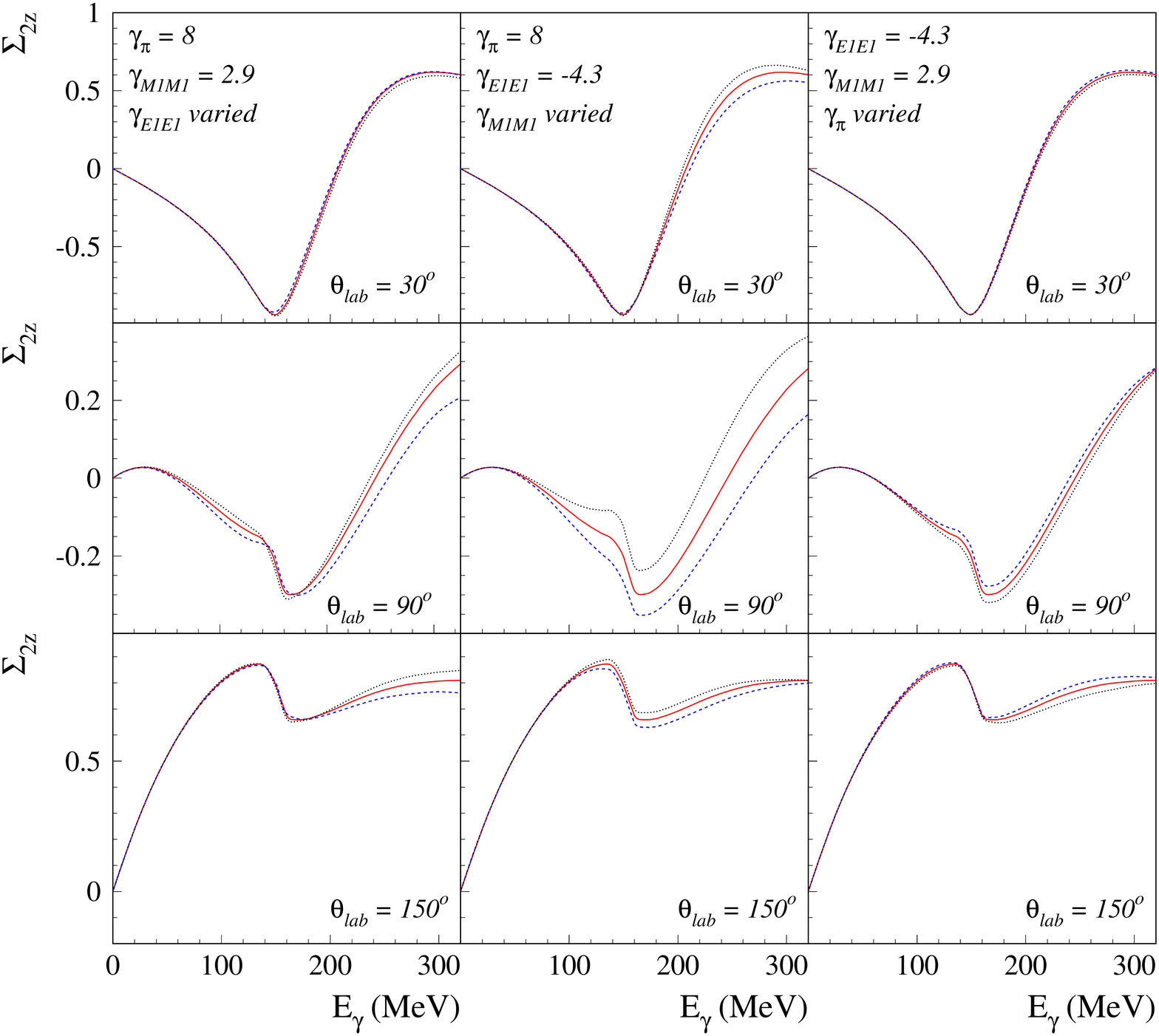,width=16 cm}
\end{center}
\caption{\small (Color online)
The beam-target asymmetry $\Sigma_{2z}$ as function of the
photon lab energy $E_{\gamma}$ plotted at different values of the photon
scattering angle $\theta_{{\rm lab}}= 30^{{\rm o }}$ (upper row), $90^{{\rm
o }}$ (central row), and $150^{{\rm o }}$ (lower row).
The results of the dispersion calculation are obtained by
using the experimental values for $\alpha_{E1}$, $\beta_{M1}$, and $\gamma_0$,
as given by Eqs.~(\ref{alpha-beta-exp}) and
(\ref{eq:go}), while the remaining
polarizabilities are taken as free parameters. Left column: results
for fixed $\gamma_{M1M1}$ and $\gamma_{\pi}$ as indicated,
and the following values of
$\gamma_{E1E1}$: $-4.3$ (red solid lines), $-2.3$ (blue dashed
lines), and $-6.3$ (black dotted lines); central column: results
for fixed $\gamma_{E1E1}$ and $\gamma_{\pi}$ as indicated,
and the following values of
$\gamma_{M1M1}$: 2.9 (red solid lines), 4.9 (blue dashed lines), and 0.9 
(black dotted lines);
right column: results for fixed $\gamma_{E1E1}$ and $\gamma_{M1M1}$ as
indicated, and the following values
of $\gamma_{\pi}$: $8$ (red solid lines), $9.8$ (blue dashed lines), and
$7.2$ (black dotted lines).}
\label{fig1}
\end{figure}

In this section we present our predictions for the polarization
observables. Our results are based on subtracted DRs evaluated with
the pion photoproduction multipoles of Refs.~\cite{HDT,said} as
input. Because of the subtraction, two-pion and heavier intermediate
states will generally yield only small corrections. Four of the
subtraction constants are determined by the experimental values for
the polarizabilities $\alpha_{E1}, \beta_{M1}, \gamma_{0},$ and
$\gamma_{\pi}$. The remaining two constants are obtained by fixing
the spin polarizabilities $\gamma_{E1E1}$ and $\gamma_{M1M1}$. In
the following figures, we start from the predictions of fixed-$t$
DRs for the dispersive part of the spin
polarizabilities~\cite{HDPV,DPV_report},
\begin{eqnarray}
\label{eq:gammaEEMM} 
\gamma_{E1E1}=-4.3 \times 10^{-4}~{\rm {fm^4}}
\,,\quad\quad \gamma_{M1M1}=2.9 \times 10^{-4}~{\rm {fm^4}} \,,
\end{eqnarray}
which are then varied by $\pm$~2 units. We note that here and in the
following discussion including the figures, the values of the spin
polarizabilities refer only to their dispersive parts, that is, the
pion pole contribution,  
$\gamma_{E1E1}^{\pi0-{\rm pole}}=-\gamma_{M1M1}^{\pi0-{\rm pole}}=
11.68\times 10^{-4}$ fm$^4$, has been subtracted. 
Because of the
relatively large error bar for $\gamma_{\pi}$, we also vary this
polarizability within its error band while keeping $\gamma_{E1E1}$
and $\gamma_{M1M1}$ fixed at their central values.
\newline
\indent Figure~\ref{fig1} shows the asymmetry $\Sigma_{2z}$, with
circular photon polarization and target aligned parallel to the
incoming photon. We observe asymmetries up to 90~$\%$ and a strong
dependence on both angle and energy, with distinct structures near
the threshold for pion photoproduction. Although the asymmetry near
$\theta_{{\rm lab}}=90^{{\rm o }}$ is smaller than for the forward
and backward directions, it is rather sensitive to a variation of
$\gamma_{M1M1}$ both near threshold and in the $\Delta~(1232)$
resonance region. Within the range of the variation, $\Sigma_{2z}$
changes by 15-20~$\%$, which provides a promising signal to
determine the spin polarizability $\gamma_{M1M1}$. The following
Fig.~\ref{fig2} displays the same information for the asymmetry
$\Sigma_{2x}$, with circular photon polarization and target aligned
sideways to the incident photon. Contrary to the previous figure,
the maximum asymmetry is now reached at scattering angles
$\theta_{{\rm lab}}\approx 90^{{\rm o }}$, and the maximum
sensitivity occurs by changing $\gamma_{E1E1}$. Within the range of
variation, this observable changes by 15~$\%$ near threshold and
40~$\%$ in the $\Delta$ region. The right panels of Figs.~\ref{fig1}
and~\ref{fig2} show that these observables are hardly changed by a
variation of $\gamma_{\pi}$. As a result, the observables
$\Sigma_{2z}$ and $\Sigma_{2x}$ sample conclusive and complementary
information on the nucleon's spin structure. Furthermore, through
the input of the DRs, they are related to the physics of the
observables E and F of pion photoproduction.
\newline
\indent The full angular distribution for $\Sigma_{2z}$ and
$\Sigma_{2x}$ are shown in Fig.~\ref{fig3} for a photon beam of
240~MeV. It is seen that the discussed sensitivity extends over a
large angular range between $30^{{\rm o }}$ and $150^{{\rm o }}$. In
view of the error bars of the backward scalar and vector
polarizabilities, $\alpha_{E1}-\beta_{M1}$ and $\gamma_{\pi}$,
respectively, the cleanest information should however come from
angles around $\theta_{{\rm lab}}\approx 90^{{\rm o }}$.
\newline
\indent Figure~\ref{fig4} displays the results for the beam
asymmetry $\Sigma_{3}$, for linear photon polarization, which is
obtained by averaging the double-polarization observable
$\Sigma_{3y}$ over the target polarizations. 
This observable rises from large negative values at low energies to 
positive values in the resonance region. Moreover, there appear interesting 
cusp effects, escpecially at forward angles. The figure shows strong
sensitivity to changes of $\gamma_{M1M1}$.  
Within the range of our variation, the asymmetry $\Sigma_3$ changes 
by about 0.15 near threshold and nearly 0.30 in the resonance region. 
The LEGS data~\cite{Blanpied01} taken at $65^{{\rm o }}$ and $90^{{\rm o }}$
are compatible with our central value for $\gamma_{M1M1}$ (full
lines), they scatter about this value at $135^{{\rm o }}$. As in the
previous examples, the variation of $\gamma_{\pi}$ leads to much
smaller effects. The following Fig.~\ref{fig5} shows the full
angular distribution for $\Sigma_{3}$ and the double-polarization
observable $\Sigma_{3y}$ in the resonance region. Large asymmetries
are predicted for both observables over a wide angular range.
Moreover, the sensitivity to the variation is large, and while
$\Sigma_{3}$ is particularly sensitive to $\gamma_{M1M1}$, the
related double-polarization asymmetry $\Sigma_{3y}$ changes strongly with
$\gamma_{E1E1}$. We conclude that these two observables are
particularly useful to disentangle the unknown spin
polarizabilities.
\newline
\indent
The physics behind the asymmetry $\Sigma_3$ is addressed in 
Fig.~\ref{fig5b} as function of the photon energy. The Born terms, 
subtraction constants, and $t$-channel contributions provide a negative 
background whose height at the resonance depends strongly  
on the value of $\gamma_{M1M1}$. The S-wave multipoles in the 
dispersion integral provide the interesting cusp effect near threshold 
through the opening of the imaginary part, whereas the P-waves are 
responsible for the further increase to large positive values in the 
$\Delta(1232)$ resonance region. In conclusion, the large effect is 
mainly given by the $M_{1+}^{(3/2)}$ multipole of pion photoproduction. 
\newline
\indent 
Let us now address the question of the model dependence of the 
dispersive approach. The general answer is that the errors in 
fixed-$t$ dispersion relations increase with 
both the beam energy and the scattering angle. 
If the energy gets larger, the unknown contributions of the higher resonances 
and backgrounds become more and more important for the evaluation of the 
$s$-channel dispersion integrals. Large angles and energies, on the other
hand, require the knowledge of the subtraction functions at large (negative) 
$t$-values. Of course, the error also depends on the specific variable under
discussion. We have studied these error sources for all the shown variables 
and found them to yield only small corrections in all the interesting cases, 
that is for large asymmetries and large sensitivity to the variation. 
Specifically for $\Sigma_3$ and $\Sigma_{3y}$ in Fig.~\ref{fig5}, 
the neglect of the two-pion production changes these asymmetries by less 
than 0.01 at all energies and angles. It is therefore reasonable to assume 
that also the neglected higher resonance and background contributions above 
$\nu = 1.5$~GeV are irrelevant for our discussion. Concerning the error from 
the $t$-channel integral, we find that the contribution from the negative 
$t$-channel cut is negligible for $E_\gamma \leq 230$~MeV. At the 
energy $E_\gamma = 275$~MeV shown in Fig.~\ref{fig5}, the asymmetry 
$\Sigma_3$ increases by less than 0.02 if we completely neglect that 
contribution, while $\Sigma_{3y}$ is even less affected by the 
$t$-channel integral. In conclusion, even a very conservative estimate yields 
errors of a few percent at most, which are almost negligible in view of the 
large range of variation predicted for $\Sigma_3$ and $\Sigma_{3y}$.  
\newline
\indent
Let us finally discuss the beam-target asymmetries with
photons linearly polarized at azimuthal angles $\phi= \pm 45^{{\rm o
}}$ with respect to the scattering plane. These asymmetries are
related to the observables G and H of pion photoproduction. They
vanish below the one-pion threshold, because they thrive on the
imaginary part of the scattering amplitudes. For the same reason,
Fig.~\ref{fig6} does not show a peak near the $\Delta$ resonance,
because all $\Delta$ multipoles carry the same phase. Instead, these
observables are strongly enhanced by interference of non-resonant
S-wave and resonant P-wave pion production. Since the pion
photoproduction multipoles are quite well known in the $\Delta$
region, we expect a rather model-independent information. The
asymmetry $\Sigma_{1z}$ turns out to be small at all angles and
energies. Figure~\ref{fig6} shows the most promising kinematics at
angles near $\theta_{{\rm lab}}=90^{{\rm o }}$, where the asymmetry
changes up to about 8~$\%$ over the range of variation. The
asymmetry $\Sigma_{1x}$, on the other hand, has its maximum
sensitivity at the backward angles. However, the variations of
$\gamma_{E1E1}$ and $\gamma_{M1M1}$ yield similar effects, such that
the observed change is qualitatively proportional to the difference
of the two polarizabilities. In conclusion, these asymmetries 
$\Sigma_{1x}$ and $\Sigma_{1z}$ would 
have to be measured with an accuracy of a few percent in order to
get useful information on the spin polarizabilities.

\section{Summary and conclusions}
\label{conclusions} Compton scattering probes the response of the
nucleon to an external electromagnetic field. At low energies, this
response is described by 2 scalar and 4 vector polarizabilities,
which contain global information on the excitation spectrum. Whereas
the scalar polarizabilities are now known with relatively small
error bars, our knowledge on the spin-dependent sector is as yet
incomplete. Only two combinations of the vector polarizabilities
have been measured. The forward spin polarizability has been
determined by forward dispersion relations as an energy-weighted
integral over the helicity-dependent total cross sections.
Furthermore, the backward spin polarizability has recently been
measured both below and in the $\Delta(1232)$ resonance region,
however with a larger model dependence. It has been known for some
time that it will take a full-fledged experimental program including
several polarization observables in order to achieve a complete
separation of the spin polarizabilities. Considerable theoretical
work has been dedicated to finding the appropriate observables and
kinematics for this purpose.
\newline
\indent In our present work we study polarized Compton scattering
within the framework of dispersion relations at $t$=const.
Subtraction of these relations speeds up the convergence of the
dispersion integrals, and the subtraction constants of the 6
relativistic amplitudes turn out to be linear combinations of the 6
polarizabilities. In most cases also the unsubtracted integrals
converge and hence a prediction for the polarizabilities is given.
However, we choose to subtract all 6 integrals in order to reduce
the model dependence on the high-energy spectrum as much as
possible. Four of the subtraction constants are then fixed by the
experimental values for the electric ($\alpha_{E1}$) and magnetic
($\beta_{M1}$) scalar polarizabilities as well as the forward
($\gamma_{0}$) and backward ($\gamma_{\pi}$) spin polarizabilities.
The remaining two independent polarizabilities are taken to be
$\gamma_{E1E1}$ and $\gamma_{M1M1}$. They are varied about a central
value as predicted by dispersion relations, within a range which
comprises several predictions from effective field theories and
dispersive approaches.
\newline
\indent Whereas the scalar polarizabilities contribute to the
differential cross section already at second order in the photon
beam energy, the vector polarizabilities appear only at third or
fourth order, depending on the observable. It is therefore not
surprising that it takes at least 100~MeV to get visible effects of
the vector polarizabilities. Even at energies near the pion
production threshold, the asymmetries change at most by $10~\%$ over
the range of our variation. Of course, the experiments could
eventually yield polarizabilities far outside the range of the
existing predictions. However we choose to stay on the conservative
side regarding our predictions. Much larger effects are obtained in
the first resonance region, in which both the cross sections and the
asymmetries increase considerably. Specifically, we find a good
chance to solve the open questions by the following two independent
setups:
\begin{itemize}
\item{}
Circularly polarized photons and targets aligned in the beam
direction ($\Sigma_{2z}$) or transverse to the beam in the
scattering plane ($\Sigma_{2x}$). Both asymmetries should yield a
reasonably large sensitivity to the spin polarizabilities in the
threshold region and increasingly large effects closer to the
resonance region. Moreover, the response to the variation shows
interesting cusp effects at threshold and a distinctly different
sensitivity of these observables with regard to $\gamma_{E1E1}$ and
$\gamma_{M1M1}$.
\item{}
Linearly polarized photons, parallel or perpendicular to the
scattering plane, and with unpolarized targets ($\Sigma_{3}$) or
with targets polarized perpendicular to the scattering plane
($\Sigma_{3y}$). The sensitivity of these two observables is
relatively small near threshold but very large in the resonance
region. Moreover, the variations of the spin polarizabilities affect
the observables in a completely different way, 
with $\Sigma_3$ being mainly sensitive to $\gamma_{M1M1}$ whereas 
$\Sigma_{3y}$ being mainly sensitive to $\gamma_{E1E1}$. 

\end{itemize}

We further looked at the observables $\Sigma_{1z}$ and
$\Sigma_{1x}$, obtained with photons linearly polarized at an angle
of $45^{\rm o }$ against the scattering plane and target alignment
perpendicular to this plane. However, these asymmetries turn out to
be small in general, and even though the relative effect of the
variations may be large, such experiments will require asymmetry
measurements at the few percent level.
\newline
\indent With the advent of new experimental tools as polarized
targets and photon beams it will be possible to study the full spin
structure of Compton scattering. The new HIGS project~\cite{HIGS} of
a high-intensity beam with circularly polarized photons in an energy
range up to 140-160~MeV is ideally suited to perform the discussed
experiments in the threshold region. Complementary investigations
should be performed at energies closer to the first resonance
region, in which we expect much larger cross sections and
sensitivities to the spin polarizabilities. Such experiments in the first 
resonance region are planned 
using the Crystal Ball detector at MAMI~\cite{CBall}.  
We strongly believe that
only a combination of such experimental projects will provide the
"sharp knife" to extract the spin polarizabilities in an unambiguous
way. The spin polarizabilities of the nucleon are fundamental
structure constants of the nucleon, just as shape and size of this
strongly interacting many-body system, which strongly justifies the
experimental effort. Such activities are both important and timely,
they will provide stringent precision tests for the existing
predictions of effective field theories and the new results expected
from the lattice gauge community for the polarizability of the
nucleon.

\section*{Acknowledgments}

The authors are grateful to J. Ahrens, R. Miskimen, and
V. Pascalutsa for useful discussions.
This research was supported by the Deutsche Forschungsgemeinschaft (SFB443),
by the DOE grant DE-FG02-04ER41302 and DOE contract DE-AC05-06OR23177 under
which Jefferson Science Associates operates the Jefferson Laboratory,
and the EU Integrated Infrastructure Initiative Hadron Physics
Project under contract number RII3-CT-2004-506078.


\begin{figure}[ht]
\begin{center}
\epsfig{file=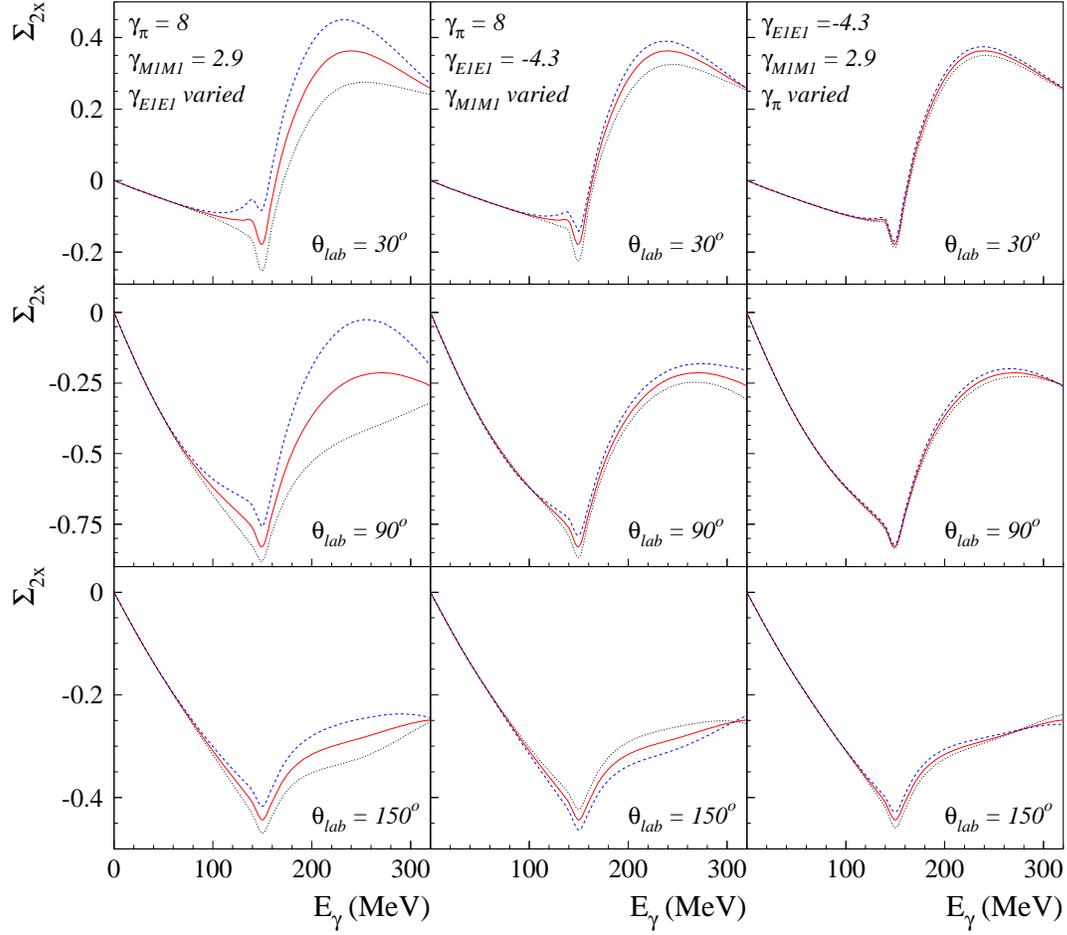,width=16 cm}
\end{center}
\caption{\small (Color online) The beam-target asymmetry $\Sigma_{2x}$ as function of the
photon lab energy $E_{\gamma}$ plotted at different values of the photon
scattering angle $\theta_{{\rm lab}}: 30^{{\rm o }}$ (upper row), $90^{{\rm
o }}$ (central row), and $150^{{\rm o }}$ (lower row). For further
notation see Fig.~\ref{fig1}.} \label{fig2}
\end{figure}
\begin{figure}[ht]
\begin{center}
\epsfig{file=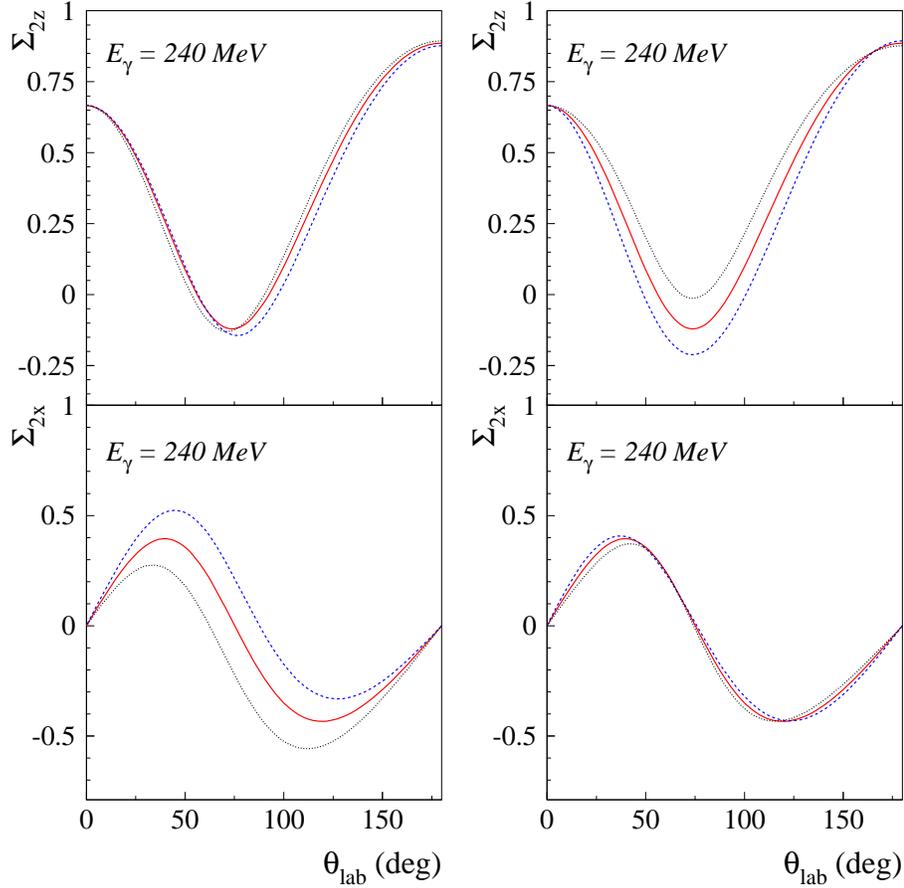,width=14 cm}
\end{center}
\caption{\small (Color online) The beam-target asymmetries
$\Sigma_{2z}$ (upper row) and $\Sigma_{2x}$ (lower row)
as function of the photon scattering  angle $\theta_{{\rm lab}}$ and at fixed photon lab energy
$E_\gamma=240$ MeV.  The results of the dispersion calculation are obtained by
using the experimental values for $\alpha_{E1}$, $\beta_{M1}$, $\gamma_0$, and
$\gamma_\pi$ as given by Eqs.~(\ref{alpha-beta-exp}), (\ref{eq:gpi}), and
(\ref{eq:go}), while $\gamma_{E1E1}$ and
$\gamma_{M1M1}$ are taken as free parameters. Left column: results
for fixed $\gamma_{M1M1}=2.9$ and the following values of
$\gamma_{E1E1}$: $-4.3$ (red solid lines), $-6.3$ (black dotted lines), and $-2.3$ (blue dashed
lines); right column: results for fixed $\gamma_{E1E1}$=$-4.3$, and the following values
of $\gamma_{M1M1}$: 2.9 (red solid lines), 4.9 (blue dashed lines), and 0.9 (black dotted lines).} \label{fig3}
\end{figure}
\begin{figure}[ht]
\begin{center}
\epsfig{file=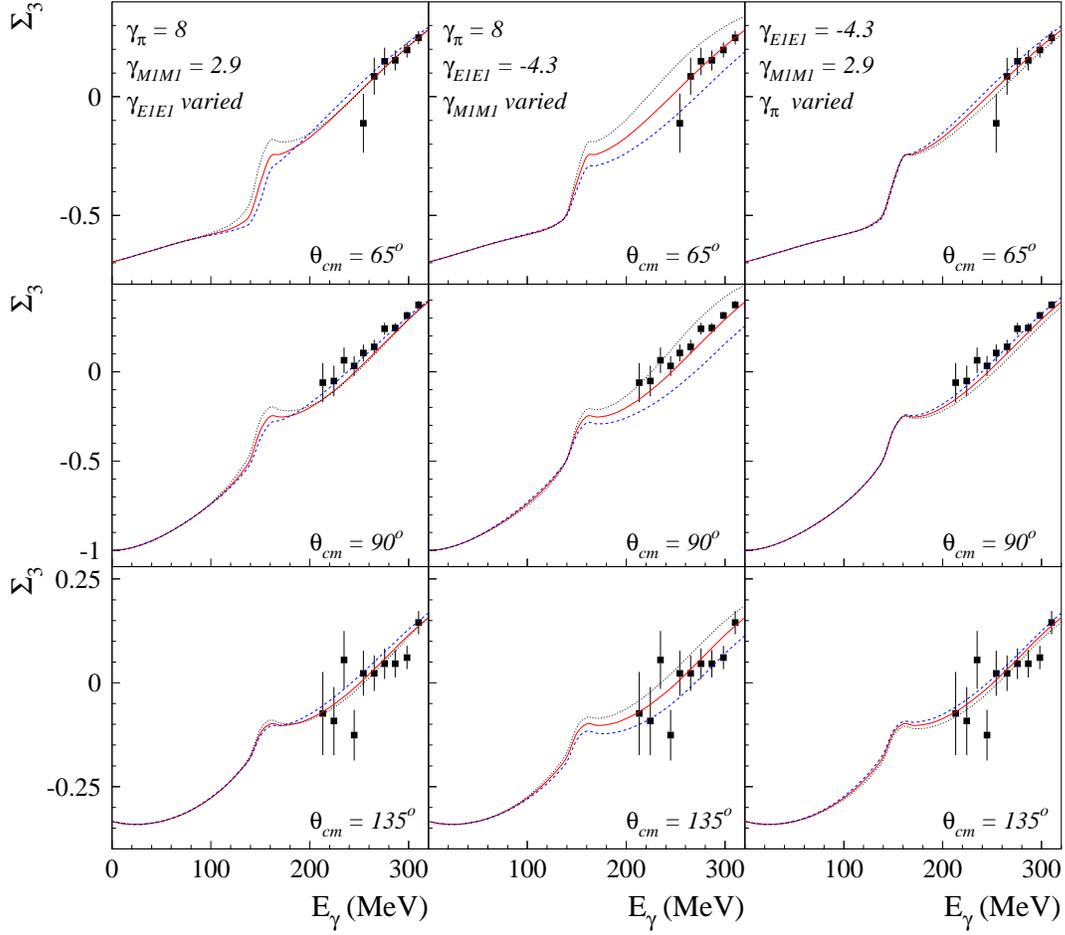,width=16 cm}
\end{center}
\caption{\small (Color online) The beam asymmetry $\Sigma_{3}$ as function of the
photon lab energy $E_{\gamma}$ plotted at different values of the photon
scattering angle $\theta_{{\rm cm}}: 65^{{\rm o }}$ (upper row), $90^{{\rm
o }}$ (central row), and $135^{{\rm o }}$ (lower row). The experimental data
are from Ref.~\cite{Blanpied01}. For further notation see Fig.~\ref{fig1}.} \label{fig4}
\end{figure}
\begin{figure}[ht]
\begin{center}
\epsfig{file=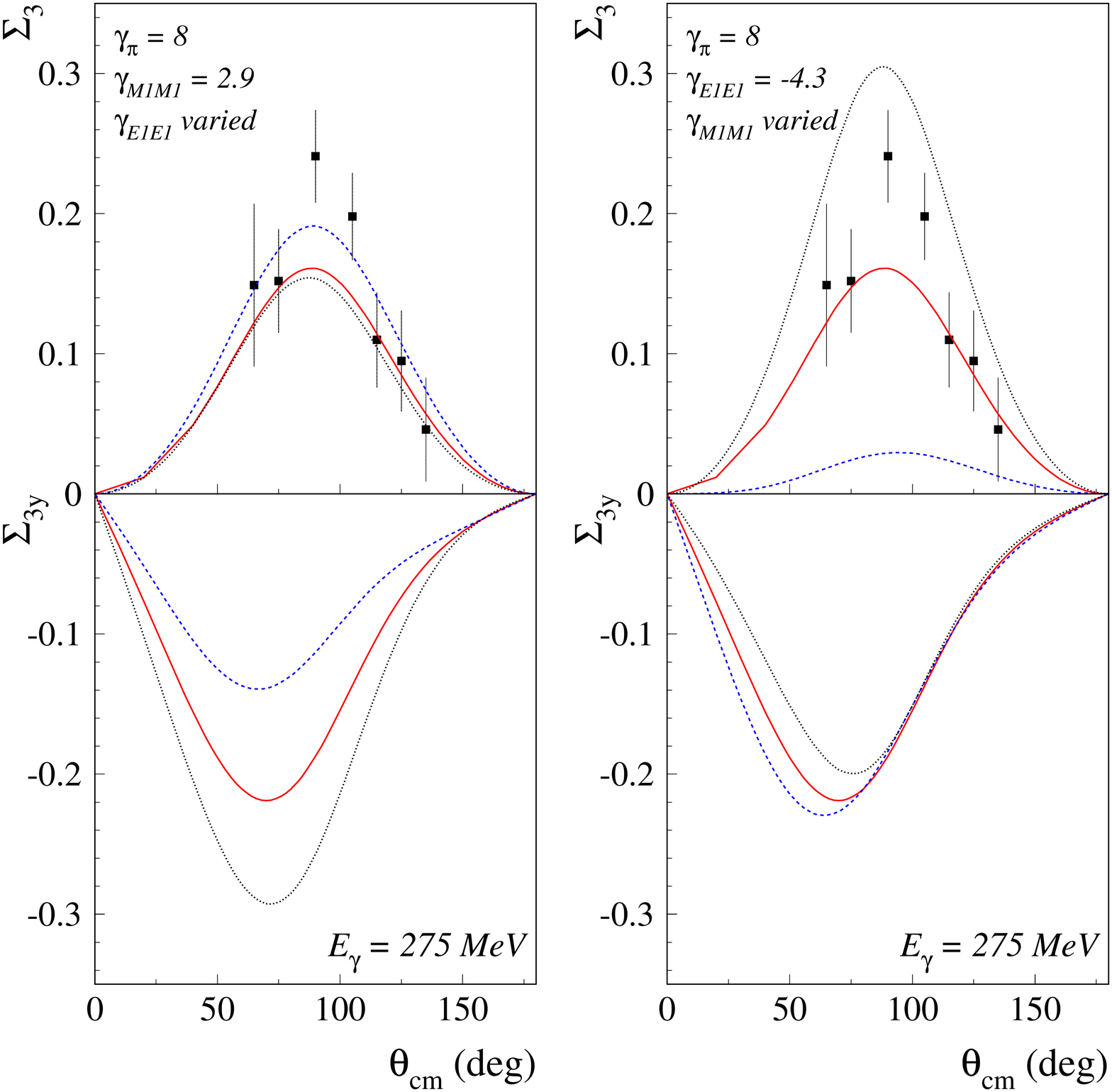,width=14 cm}
\end{center}
\caption{\small  (Color online) The  beam asymmetry $\Sigma_{3}$ (upper row) 
and beam-target asymmetry
$\Sigma_{3y}$ (lower row) as function of the photon scattering  angle $\theta_{{\rm cm}}$
and at fixed photon lab energy $E_\gamma=275$ MeV. The experimental data
are from Ref.~\cite{Blanpied01}. For further notation see Fig.~\ref{fig3}.}
\label{fig5}
\end{figure}
\begin{figure}[ht]
\begin{center}
\epsfig{file=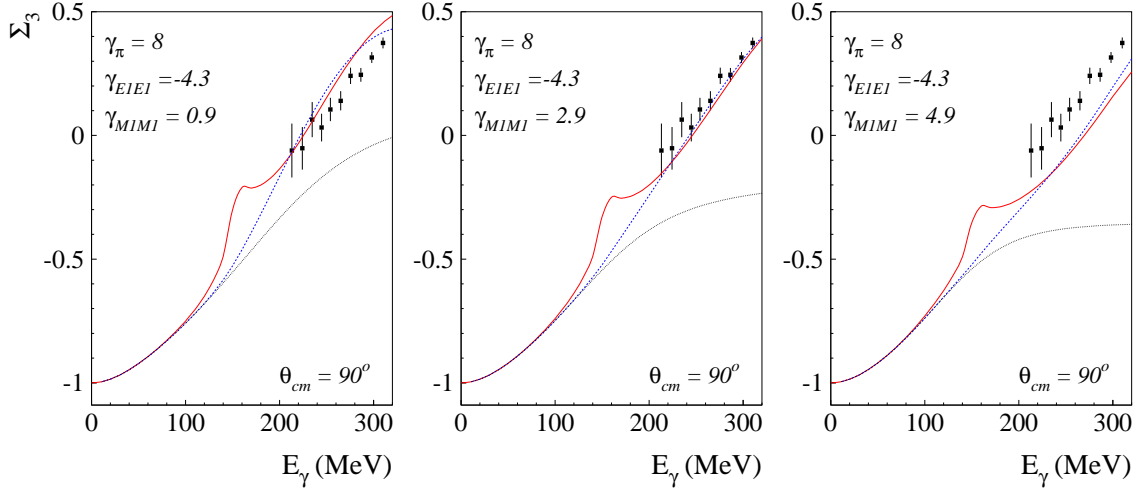,width=16 cm}
\end{center}
\caption{\small (Color online) Different contributions in the dispersion 
relation calculation to the beam asymmetry $\Sigma_{3}$ as function of the
photon lab energy $E_{\gamma}$ at photon scattering angle of 
$90^{{\rm o}}$, for different values of $\gamma_{M1M1}$ as indicated 
on the figure. The black dotted curves show the contribution from Born diagrams + 
subtraction constants + $t$-channel. The blue dashed curves are the contribution 
from Born diagrams + subtraction constants + $t$-channel + $M_{1+}$ 
multipole in the $s$-channel. The red solid curves are the total result. 
The experimental data are from Ref.~\cite{Blanpied01}.} 
\label{fig5b}
\end{figure}
\begin{figure}[ht]
\begin{center}
\epsfig{file=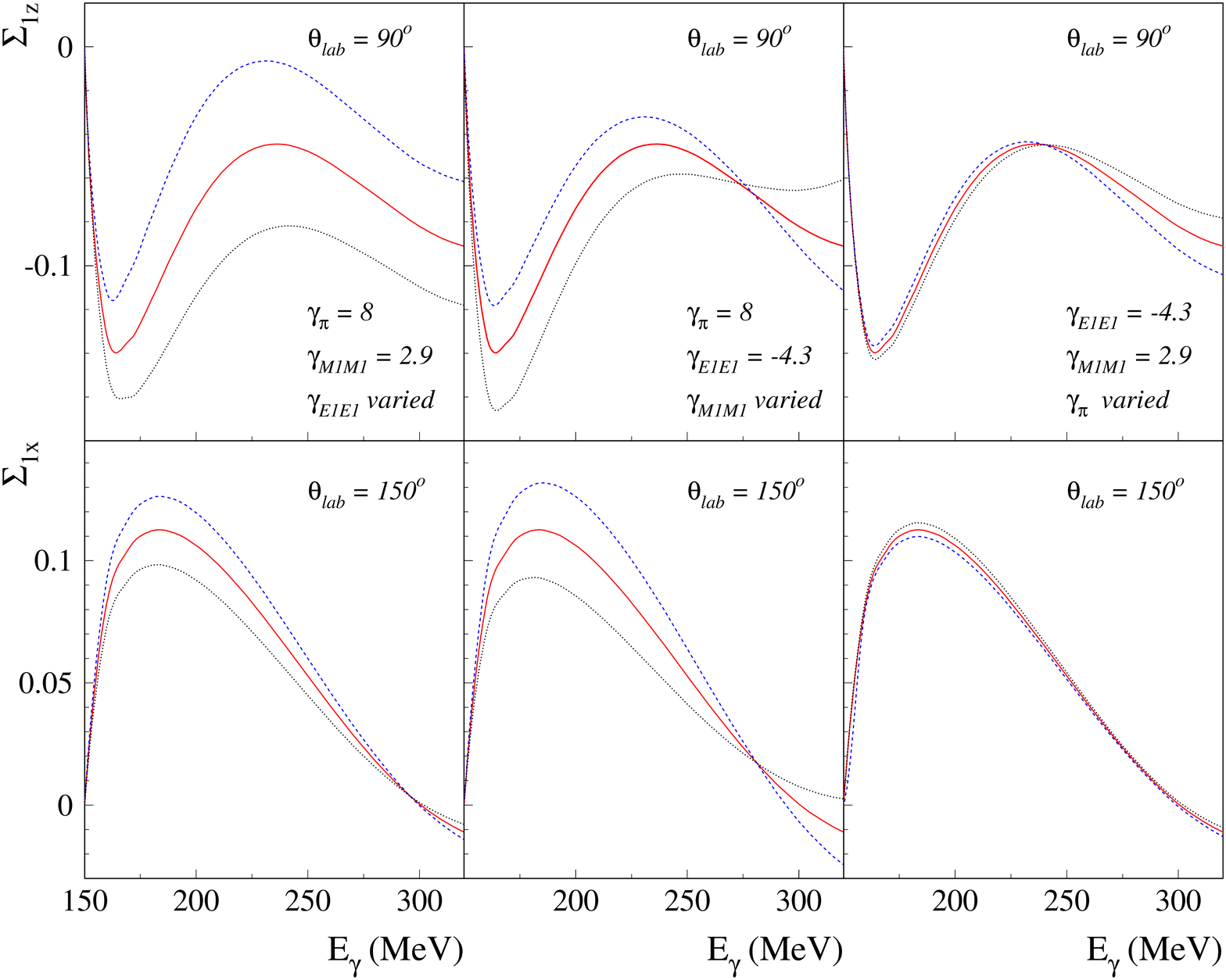,width=16 cm}
\end{center}
\caption{\small  (Color online)
The beam-target asymmetries $\Sigma_{1z}$ (upper row) at photon scattering angle
$\theta_{{\rm lab}}$=$90^{{\rm o }}$ and $\Sigma_{1x}$ (lower row) at
$\theta_{{\rm lab}}$=$150^{{\rm o }}$, plotted as function of the photon
lab energy $E_{\gamma}$. Further notation as in Fig.~\ref{fig1}.} \label{fig6}
\end{figure}
\end{document}